\documentclass[aps,prl,twocolumn,showpacs]{revtex4}
\usepackage{graphicx}
\usepackage{amssymb}
\usepackage{color}
\begin{document}

%\preprint{}

\title{Series expansion analysis of a tetrahedral cluster spin chain}



\author{Marcelo Arlego and Wolfram Brenig}

\address{Institut f\"ur Theoretische Physik, Technische
Universit\"at Braunschweig, 38106 Braunschweig, Germany}

\email[E-mail: ]{m.arlego@tu-bs.de}


\date{\today}

\begin{abstract} Using series expansion by continuous unitary
transformations we study the magnetic properties of a frustrated tetrahedral
spin$-\frac{1}{2}$ chain. Starting from the limit of isolated
tetrahedra we analyze the evolution of the ground state
energy and the elementary triplet dispersion as a function of the
inter-tetrahedral coupling.
The quantum phase diagram is evaluated and is shown to
incorporate a singlet product, a dimer, and a Haldane phase.
Comparison of our results with those from several other techniques,
such as density matrix renormalization group, exact diagonalization and
bond-operator theory are provided and convincing agreement is
found.
\end{abstract}

% insert suggested PACES numbers in braces on next line
%\pacs{PACES number(s): 45.10.-b, 45.50.Tn, 45.70.-n, 62.20.-x, 81.40.Pq}

%\keywords{}

\maketitle

%%%%%%%%%%%%%%%%%%%%%%%%%%%%%%%%%%
%\section{INTRODUCTION}
%%%%%%%%%%%%%%%%%%%%%%%%%%%%%%%%%%
\section{Introduction}

Geometric frustration is a central issue in present days research
on quantum magnets. In particular, materials with spin-tetrahedra
as building blocks are under scrutiny, since they bear the
potential of exotic magnetic properties \cite{DiepBook2005}. Many
investigations have focused on dense tetrahedral magnets, eg. on
pyrochlore lattices \cite{DiepBook2005}. Quantum magnets with
spatially more separated spin-tetrahedra have been considered far
less exhaustive. The isostructural tellurate compounds
Cu$_2$Te$_2$O$_5$X$_2$ (X=Br or Cl) \cite{Johnsson00,Lemmens01}
are a prominent example of such `diluted' systems. They contain
disjoint tetrahedral clusters of Cu$^{2+}$ with $S=1/2$
\cite{Johnsson00}. The tetrahedra display two longer and four
shorter edges. They form chains along the $c$ direction and are
separated by lone-pair cations in the $a,b$ plane
\cite{Johnsson00}. Analysis of the susceptibility
\cite{Johnsson00,Gros03,Gros03b} is consistent with a strongly
frustrated antiferromagnetic {\em intra}-tetrahedral exchange of
$J_1=47.5(40.7)K$ along the shorter edges and a ratio of
$J_2/J_1=0.7(1)$ between the long and the short edges for
X=Br(Cl). Both systems show incommensurate ordering with strongly
reduced moments \cite{Zaharko04} below a transition temperature of
$T_c=11.4(18.2)K$ for X=Br(Cl). This $T_c$ is consistent with bulk
thermodynamics \cite{Johnsson00,Lemmens01,Gros03,Prester04}.
%Lattice degrees of
%freedom may be involved in this transition \cite{Prester04}.
Observation of longitudinal magnons suggests that
the Br-system could be close to quantum criticality
\cite{Lemmens2002a,Gros03}.

Despite the low-temperature ordering of the tellurates, the
effects of {\em inter}-tetrahedral coupling at various
dimensionality and on an energy scale of the intra-tetrahedral
exchange are open issues in such magnets. In the tellurates
dispersive features have been observed in inelastic neutron
scattering on polycrystals \cite{Crowe2005a}. Two-magnon Raman
scattering provides evidence for inter-tetrahedral dispersion
along the $c$-axis only \cite{Lemmens01}. Considerations of the
electronic structure however, suggested that $a,b$-planar exchange
via O-Te-O bonds is important also \cite{Valentí03,Whangbo2003a}.

Theoretically, inter-tetrahedral coupling in diluted tetrahedral
magnets has been analyzed in terms of one-dimensional (1D)
tetrahedral spin chains using bond-operator methods, exact
diagonalization \cite{Brenig-tet} and effective Hamiltonians
\cite{Mikeska02}. The chains have been found to allow for a
singlet, a dimer, and a Haldane phase with low-energy singlet
excitations close to the dimer-to-singlet transition which is of
first order. In 3D molecular field theories \cite{Gros03,Gros03b}
have been used to model the magnetic transition in the tellurates.
Moreover series expansion (SE) up to 4th order in various 3D
inter-tetrahedral couplings has been carried out
\cite{Brenig-tet03} resulting in a rich quantum phase diagram with
several competing states. Motivated by this situation, the aim of
the present work is to extent and improve the analysis of 1D
tetrahedral chains carried out in ref. \cite{Brenig-tet} by
employing the SE method of ref. \cite{Brenig-tet03}, however in 1D
only, which will enable us obtain results to higher orders.

\begin{figure}[tb]
\begin{center}
\includegraphics[width=0.45\textwidth]{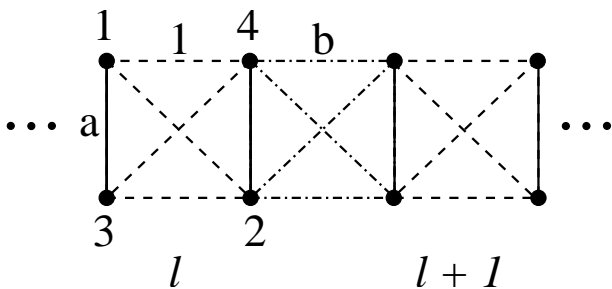}
\end{center}
\caption{Tetrahedral cluster-chain. The unit cell, labelled with
$l$, contains spin$-1/2$ moments ${\bf s}_{i\,l}$ at vertices
$i=1,\ldots,4$ (solid circles). All couplings are in units of
$J_1$ (dashed lines).} \label{chain}
\end{figure}

Fig.\ref{chain} depicts the tetrahedral spin$-1/2$ chain
introduced in refs. \cite{Brenig-tet,Mikeska02}. The tetrahedra
correspond to the sites 1-4 with couplings $J_1$ and $J_2=a J_1$.
Inter-tetrahedral exchange occurs through $J_3=b J_1$. The Hilbert
space of a single tetrahedron consists of 16 states, i.e., two
singlets ${\cal S}_{1,2}$, three triplets ${\cal T}_{1,2,3}$ and
one quintet ${\cal Q}$ whose energies are listed in
Table~\ref{tab}. This level scheme implies that a singlet resides
within the singlet-triplet gap of the tetrahedron for $1/2<a<2$.
The Hamiltonian of the chain can be written in terms of the total
edge-spin operators ${\bf P}_{1(2),l}={\bf s}_{1(4),l}+{\bf
s}_{3(2),l}$, where ${\bf s}_{i,l}$ denotes a spin$-1/2$ at vertex
$i$ on the tetrahedron at site $l$. For the $P_i$ quantum numbers
refer to Table~\ref{tab}.
\begin{eqnarray}
\frac{H}{J_1}= \sum_{l=1}^{L/2} [{\bf P}_{1,l}{\bf P}_{2,l} +
\frac{a}{2} ({\bf P}^2_{1,l} + {\bf P}^2_{2,l} -3)+b {\bf
P}_{2,l}{\bf P}_{1,l+1}]. \label{bb1}
\end{eqnarray}
Here $L$ is the number of rungs (pairs of sites coupled by $J_2=a
J_1$). The Hamiltonian commutes with the edge spin, i.e. $[H,{\bf
P}^2_{i(=1,2),l}]=0;\,\forall\,l,i=1,2$. Hence, the Hilbert space
decomposes into sectors of fixed distributions of locally
conserved edge-spin eigenvalues $P_{i,l}$, each corresponding to a
sequences of spin$-1$ chain-segments ($P_{i,l}=1$) intermitted by
chain-segments of {\em localized} singlets ($P_{i,l}=0$).
It has been shown in ref. \cite{Brenig-tet} that the ground state
of Eq.(\ref{bb1}) occurs only within the homogeneous sector of
purely $\mathcal{S}_2$ states or in the $P_{i,l}=1,\, \forall i,l$
\, sector. I.e., inhomogeneous phases consisting of both
$P_{i\,l}=0$ {\em and} $P_{i\,l}=1$ sites will not be ground
states. Keeping $b$ fixed and for $a\rightarrow\infty$, the ground
state must result from local ${\cal S}_2$ type of states. This is
a trivial, decoupled \emph{singlet product} phase with singlets on
each rung. The ground state energy per tetrahedron in this phase
is independent of $b$ and it is given by: $e_{g\,\mathcal{S}_2}=
-3 a / 2$. In the opposite limit $a \rightarrow 0$ the ground
state is in the $P_{i,l}=1,\, \forall i,l$\, sector. There is a
line of a first order quantum phase transition in the
$(a,b)$-plane that switches between both ground states, beginning
at the non-interacting point $a=1, b=0$. In the next Sections we
will determine this line by SE.

In the $P_{i,l}=1,\, \forall i,l$\, sector the tetrahedral chain
is equivalent to that of a dimerized spin$-1$ chain with $L$ sites
\begin{eqnarray}
\frac{H}{J_1}= \sum_{l=1}^{L}[{\bf S}_{2l-1}{\bf S}_{2l}+b {\bf
S}_{2l}{\bf S}_{2l+1}]+\frac{a}{4}L, \label{w3}
\end{eqnarray}
where the edge-spin operators
$\mathbf{P}_{1,l}=\mathbf{S}_{2l-1}$\, and
$\mathbf{P}_{2,l}=\mathbf{S}_{2l}$ have been relabelled in terms
of spin$-1$ operators $\mathbf{S}_l$. For $b=0$, the chain
decomposes in isolated dimers and for $b=1$\, it is the
homogeneous spin$-1$ chain. In both cases the system has a spin
gap. Trivially in the former one and according to Haldane
\cite{Haldane1983-1,Haldane1983-2} in the latter.  However, for a
particular $b_c$ the gap closes, defining a second order quantum
phase transition between the \emph{dimer phase} $(b < b_c)$\, and
the \emph{Haldane phase} $(b > b_c)$. This transition will be
analyzed in the next Sections by calculating the evolution of the
gap in terms of the perturbative parameter $b$ using SE around the
non-interacting point $(b=0)$. Our results will be compared with
those obtained by other methods.

\begin{table}[ht]
\begin{tabular}{c|r|r|r}
\hline \hline & $P_1$ & $P_2$ & $E/J_1$ \\ \hline
${\cal S}_1$     & 1 & 1 & $-2+a/2$ \\
${\cal S}_2$     & 0 & 0 & $-3a/2$ \\
${\cal T}_1$     & 1 & 1 & $-1+a/2$ \\
${\cal T}_{2,3}$ & 0,1 & 1,0 & $-a/2$ \\
${\cal Q}$       & 1 & 1 & $1+a/2$ \\
\hline \hline
\end{tabular}
\caption{Eigenstates of a single tetrahedron: singlets $({\cal
S}_1, {\cal S}_2 )$; triplets $({\cal T}_1, {\cal T}_{2,3})$ and
Quintet $({\cal Q})$. Each state is labelled by the $P_{1,2}$
edge-spin quantum numbers and the energy $E/J_1$. Site index $l$
suppressed.} \label{tab}
\end{table}

\section{Series expansion by continuous unitary transformation}
\ The Hamiltonian of the tetrahedral chain (Eq.(\ref{bb1})) can be
written as
\begin{equation}\label{tet-red}
    \frac{H}{J_{1}} = H_{0}+ a T_{a\,0} + b\sum_{n=-N}^{N} T_{b\,n}.
\end{equation}
$H_0$ is the sum over local tetrahedral Hamiltonians at $a=0$.
Their spectra consist of four equidistant energy levels $E/J_1 =
-2, -1, 0, 1$  (see Table \ref{tab}). With these levels we
associate a number $q_l$ of local energy \emph{quanta} $q_l = 0,
\ldots, 3$. Together with $P_{1,2}$, this characterizes the basis.
$H_0$ has an equidistant ladder spectrum labelled by $Q = \sum_l
q_l$. $Q=0$ refers to the \emph{unperturbed} ground state of
$H_0$: $|0\rangle \equiv |\mathcal{S}_{1,1}\ldots
\mathcal{S}_{1,L/2} \rangle$, i.e. an $\mathcal{S}_1$ singlet
product. The $Q=1$ sector of $H_0$ consists of linear combinations
of local $\mathcal{T}_{1,j}$ triplet excitations $|t_j\rangle
\equiv |\mathcal{S}_{1,1}\ldots \mathcal{T}_{1,j} \ldots
\mathcal{S}_{1,L/2} \rangle$, with a $\mathcal{T}_{1,j}$ triplet
on the tetrahedron at site $j$. $T_{a\,0}$ refers to the sum over
the local terms proportional to $a$ in Eq.(\ref{bb1}). By
construction this term is diagonal in the basis of $H_0$. The
third term in Eq.(\ref{tet-red}) refers to the inter-tetrahedral
coupling via $b$. The operators $T_{b\,n}$ non-locally create
(destroy) $n \geq  (<)\, 0$ quanta within the ladder spectrum of
$H_0$. For our model $N \leq N_{max}\equiv 6$, in principle.
Explicit calculation of the $T_{bn}$  however shows that $N \leq
4$ \cite{electronic_data}.

Note that by a shift of one half of the unit cell, i.e.~${\bf
P}_{2\,l(1\,l+1)}\rightarrow {\bf P}_{1\,l(2\,l)}$  Eq.(\ref{bb1})
is symmetric under the operation $(J_1,a,b)\rightarrow (J_1
b,a/b,1/b)$. Therefore, in order to cover the {\em complete}
parameter space for $a,b>0$ it is sufficient to consider the phase
diagram in the range of $a\in[0,\infty]$ and $b\in[0,1]$.

It has been shown \cite{Stein97,Mielke98,Knetter00} that models of
type Eq.(\ref{tet-red}) allow for high-order SE using a continuous
unitary transformation generated by the flow equation method of
Wegner \cite{Wegner94}. Adapted to our case, the mapping of
Eq.(\ref{tet-red}) onto the unitarily rotated effective
Hamiltonian $H_{\rm eff}$ reads \cite{Stein97,Knetter00}
\begin{eqnarray}
\label{Heff}
 \nonumber  H_{\rm eff} &=& H_{0}+ a T_{a\,0} \\
   & & +\sum^\infty_{n=1}\; b^n
\sum_{\stackrel{\mbox{\scriptsize $|{\bf m}|=n$}}{M({\bf m})=0}}
C({\bf m})\;T_{b\,m_1}T_{b\,m_2}\ldots T_{b\,m_n}, \nonumber \\
\end{eqnarray}
where ${\bf m}=(m_1,\ldots ,m_n)$ is an $n=|{\bf m}|$-tuple of
integers, each in a range of $m_i\in\{0,\pm 1,\ldots,\pm N\}$ and
$M({\bf m})\equiv\sum^n_{i=1} m_i$. A main advantage of this
method is that in  contrast to $H$ of Eq.(\ref{tet-red}), $H_{\rm
eff}$ is constructed to \emph{conserve} the total number of quanta
$Q$ at each order $n$. This is clear from the constraint $M({\bf
m})=0$. $Q$ conservation allows the SE of several quantities using
the bare eigenstates of $H_0$ like the ground state energy and the
elementary triplet dispersion. This will be done in the next
Sections. The amplitudes $C({\bf m})$ are rational numbers
computed from the flow equation method \cite{Stein97,Knetter00}.
The $C({\bf m})$ table is available on \cite{electronic_data}.

\section{Ground state energy and elementary triplet dispersion}
Now we discuss results for the ground state energy $E_g$ and the
triplet dispersion $\omega(k)$ in the dimer phase as obtained from
SE with respect to the inter-tetrahedral coupling $b$.
$Q$-conservation leads to
\begin{equation}\label{Eg}
    E_g=\langle 0|H_{\rm eff}|0\rangle, \,
\end{equation}
where $|0\rangle = |\mathcal{S}_{1,1}\ldots \mathcal{S}_{1,L/2}
\rangle$ and $H_{\rm eff}$ is the effective Hamiltonian given by
Eq.(\ref{Heff}). Wrap-around of graphs up to length $n$ will not
occur if this matrix element is evaluated on chains with $n+1$
tetrahedral clusters and periodic boundary conditions (PBC). This
is required by linked-cluster theorem \cite{Gelfand-review} and
leads to SE's valid to $O(n)$ in the thermodynamic limit, i.e. for
infinite-sized systems.

\begin{figure}[tb]
 \begin{center}
\includegraphics[angle=270,width=0.48\textwidth]{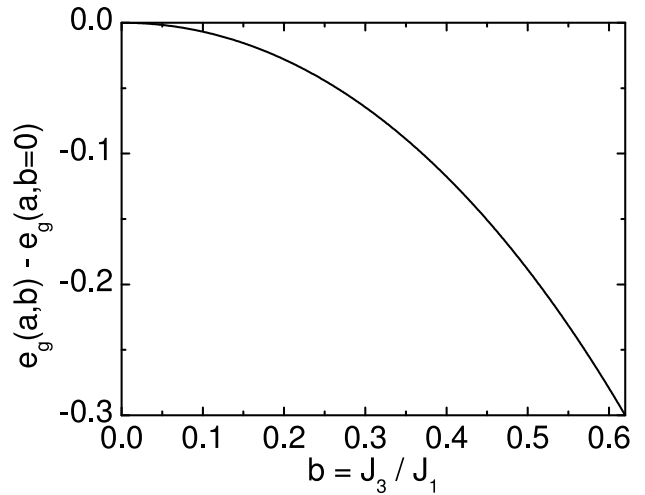}
 \end{center}
\caption{Ground state energy $e_g (a,b)-e_g (a,b=0)$ per
tetrahedron in the dimer phase.\label{egs}}
\end{figure}

We have evaluated $E_g$ up to $O(7)$. The ground state energy per
tetrahedron $e_g$ in the dimer phase reads
\begin{eqnarray}
\label{eg} e_g (a,b)=&& -2 +\frac{a}{2} - \frac{2\,b^2}{3} -
\frac{b^3}{6} +
  \frac{b^4}{108} - \frac{67\,b^5}{1620}
\nonumber\\
&& - \frac{53273\,b^6}{1749600} +
\frac{27311519\,b^7}{5038848000}.
\end{eqnarray}
The first two terms in Eq.(\ref{eg}) correspond to the
non-interacting energy of the ${\cal S}_1$ state as in Table
\ref{tab}. For $a=0$ Eq.(\ref{eg}) is also a SE of the ground
state energy per dimer of the dimerized spin$-1$ chain. To the
best of our knowledge an analytic expression of this has not been
published previously. Fig.\ref{egs} depicts the ground state
energy which is a monotonously decreasing function of the
inter-tetrahedral coupling $b$. We find, that plots of $e_g(a,b)$
to $O(6)$ and $O(7)$ are indistinguishable on the scale of
Fig.\ref{egs}, which provides an estimate of convergence. Using
$e_g(a,b)$ we will discuss the first order critical line
$b_c(a_c)$ for the dimer-to-singlet transition in the next
Section. Trivially, the non-interacting critical point is
$b_c(1)=0$ (table \ref{tab}).

To calculate the triplet dispersion we have to consider the
subspace of exactly one $\mathcal{T}_{1,j}-$type of excitation:
$|t_j\rangle = |\mathcal{S}_{1,1}\ldots \mathcal{T}_{1,j} \ldots
\mathcal{S}_{1,L/2} \rangle$, i.e. one-triplet $\mathcal{T}_{1,j}$
on a tetrahedron $j$. By $Q$ conservation the number of triplets
is conserved, implying that $H_{ \rm{eff}}$ can only translate the
triplets $|t_j\rangle $
\begin{equation}
\label{c-coeffs}
  H_{\rm{eff}}|j\rangle=\sum_i c_i|j+i\rangle \ .
\end{equation}
Due to translational invariance of our model, spatial Fourier
transformation $|k\rangle=\sqrt{4/L}\sum_j \exp(ikj)|j\rangle$,
diagonalizes Eq.(\ref{c-coeffs}) leading to the triplet dispersion
\begin{eqnarray}
\label{disp-gral}
  \omega(k)& \equiv &\left<k\right|H_{\rm{eff}}|k\rangle-E_{g} \nonumber\\
 &=&  c_0-E_{g}+ \sum_{j=1}^{\infty} 2c_j  \cos(kj) \ .
\end{eqnarray}
We have evaluated the coefficients $c_i$ up to $O(7)$. From them,
Eq.(\ref{disp-gral}) in the thermodynamic limit reads
\begin{eqnarray}
\label{dispersion}
 &&\omega (k,b) =
 \left( 1+ \frac{8 \,b^2}{27} +\frac{ 19\,b^3}{27} -\frac{571 \,b^4}{972}
 +\frac{ 183943\,b^5}{233280} \right. \nonumber\\
 && \left. -\frac{ 391390595851\,b^6}{380936908800}+\frac{913820919969227 \,b^7}{511979205427200} \right)
 \nonumber\\
 && +\left( -\frac{4\,b}{3}-\frac{2\,b^2}{3}+\frac{26\,b^3}{27}-\frac{29\,b^4}{54} +\frac{145237\,b^5}{98415} \right. \nonumber\\
 && \left.  -\frac{4087919\,b^6}{1959552}
 +\frac{10916063988776383\,b^7}{2879883030528000} \right) \, \cos(k) \nonumber\\
 && +\left( -\frac{4\,b^2}{9}-\frac{2\,b^3}{9}-\frac{7285\,b^4}{34992}+\frac{912407\,b^5}{699840}\right. \nonumber\\
 && \left.
 -\frac{18113617135\,b^6}{10158317568}+\frac{485683037077901\,b^7}{142216445952000}\right)\, \cos(2k) \nonumber\\
 && +\left(-\frac{8\,b^3}{27}-\frac{32\,b^4}{81}+\frac{1558441\,b^5}{3149280}-\frac{135853999\,b^6}{220449600}\right. \nonumber\\
 && \left.
 +\frac{53288026128863\,b^7}{21332466892800}\right)\, \cos(3k) \nonumber\\
 && +\left(-\frac{20\,b^4}{81}-\frac{268\,b^5}{729}-\frac{22011727\,b^6}{94478400}\right. \nonumber\\
 && \left.
 +\frac{5321414051\,b^7}{3429216000}\right)\, \cos(4k) \nonumber\\
 && +\left(-\frac{56\,b^5}{243}-\frac{3068\,b^6}{6561}+\frac{2345836457\,b^7}{25509168000}\right)\, \cos(5k) \nonumber\\
 && +\left(-\frac{56\,b^6}{243}-\frac{32332\,b^7}{59049}\right)\, \cos(6k) \nonumber\\
 && +\left(-\frac{176\,b^7}{729}\right)\, \cos(7k). \nonumber \\
\end{eqnarray}
As for $e_g$ and since $\omega (k,b)$ is independent of $a$, this
result may also be interpreted as a SE for the triplet dispersion
of a dimerized spin$-1$ chain.

\begin{figure}[tb]
\begin{center}
\includegraphics[angle=-90,width=0.48\textwidth ]{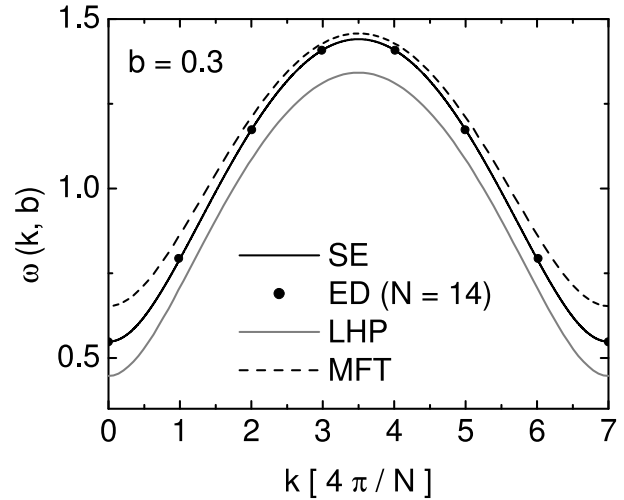}
\end{center}
\caption{Elementary triplet dispersion in the spin$-1$ dimer phase
for $b=0.3$. Solid line: SE from Eq.(\ref{dispersion}). Gray line,
dashed line, and solid dots: bond-boson Mean Field Theory (MFT),
bond-boson Linear Holstein Primakoff (LHP) approximation, and
exact diagonalization (ED) from ref. \cite{Brenig-tet}.}
\label{disp}
\end{figure}

In Fig.\ref{disp}\, we compare our SE results (solid lines) for
$\omega(k,b)$ with other findings at an intermediate value of $b$
which will be shown later to reside still within the dimer phase.
The first excited eigenvalues obtained from exact diagonalization
(ED) on a finite dimerized spin$-1$ chain \cite{Brenig-tet} are
displayed with dots. Results from bond-boson mean field theory
(MFT) \cite{Brenig-tet} and linearized Holstein-Primakoff (LHF)
methods \cite{Brenig-tet,Starykh96a,eder98} are shown with gray
and dashed lines, respectively. As we can observe in this Fig. the
agreement between the SE's results and ED is very good, and the SE
clearly improves on earlier uncontrolled approximations.
Eq.(\ref{dispersion}) shows that there is a tendency to the
closure of the gap: $\Delta \equiv \omega(k=0,b)$ as $b$
increases, which signals the transition to the Haldane phase. This
will be analyzed in detail in the next Section. Finally we note,
that in the 1D case our results extend to $O(7)$ an earlier SE
which was obtained to $O(4)$ in ref. \cite{Brenig-tet03}, however
for a 3D tetrahedral spin model.

\section{Quantum phase transitions}
\begin{figure}[tb]
 \begin{center}
 \includegraphics[angle=270,width=0.48\textwidth]{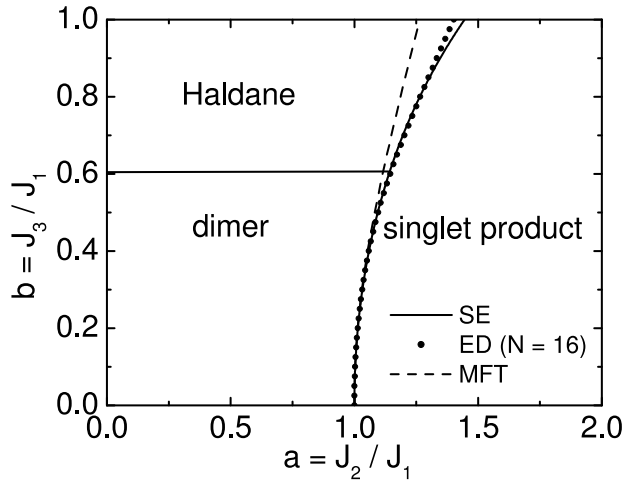}
 \end{center}
\caption{Quantum phase diagram of the tetrahedral cluster chain.
Solid: first order dimer-to-singlet critical line $b_c(a_c)$
obtained from SE. Dashed and dots: $b_c(a_c)$ from bond-boson mean
field theory (MFT) and exact diagonalization (ED) in the dimerized
spin$-1$ sector for $N=16$ sites and PBC (ref. \cite{Brenig-tet}).
Horizontal solid line at $b_c \in [0.612,\, 0.614]$: second order
dimer-to-Haldane transition obtained by Dlog-Pad\'e analysis of
the triplet-gap closure of Eq.(\ref{dispersion}). (See also
Fig.\ref{gap}).} \label{phases}
\end{figure}

In this Section we consider the quantum critical points of our
model. Fig.\ref{phases} summarizes the phase diagram of the
tetrahedral cluster spin chain as determined from the present SE
and other methods \cite{Brenig-tet,Mikeska02}. It displays the
singlet product, dimer and Haldane phases. These regions are
separated by a first order critical line between the dimerized
spin$-1$ chain sector and the singlet product phase, as well as
the second order critical line between the dimer and the Haldane
phases.

We start by studying the dimer-to-singlet first order transition.
The solid line represents our evaluation of the critical line
$b_c(a_c)$  using SE. It has been obtained by numerical
determination of $b(a)$ from $ e_g(a,b)- e_{g\,\mathcal{S}_2} =
0$, where $e_{g\,\mathcal{S}_2}= -3 a / 2$ is the exact ground
state energy of the singlet-product phase. Obviously, there is an
excellent agreement  with ED results \cite{Brenig-tet,Golinelli94}
(dotted line in Fig.\ref{phases}). Surprisingly this is true even
for $b$-values beyond the dimer-Haldane transition, where the SE
is not expected to be valid anymore. From this we conclude that
the ground state energy of the Haldane phase has little difference
to that of an adiabatically continued dimer phase. While the
accuracy of approximate approaches, as eg. bond-operator MFT
\cite{Brenig-tet}, is hard to asses, it is interesting to see that
the SE agrees well also with the latter approach at least up to
the dimer-Haldane transition (dashed line in Fig. \ref{phases}).

The dimer-Haldane transition of the dimerized spin$-1$ chain has been
analyzed by several authors. From the analytical point of view,
Haldane and Affleck \cite{Affleck-Haldane87, Affleck85} have performed
a mapping onto the O$(3)$ nonlinear $\sigma-$model (NLSM) in the large
$S$ limit. They have shown that the \emph{topological angle} $\theta$
is given by $\theta=2\pi S(1-\delta)$ and that the system is gapless
and described by a conformal field theory with SU$(2)$ symmetry when
$\theta/(2 \pi)$ is half odd integer. $\delta$ is related  to $b$ by
$b=(1-\delta)/(1+\delta)$. Therefore, for $S=1$ a gapless point was
predicted to occur at $b_c=1/3$, i.e. $\delta_c=1/2$. While the prediction
of a gapless point from the NLSM is expected to be
correct, quantitative agreement for the critical point is not to be
expected since $1/S$-correction may play a role and have not been
analyzed to our knowledge.

From the numerical point of view, the dimer-Haldane transition has
been studied employing Density Matrix Renormalization Group (DMRG)
\cite{Kato94,Zheludev04}, Quantum Monte Carlo (QMC)
\cite{Yamamoto95, Kohno98}, and ED techniques
\cite{Brenig-tet,Totsuka95,Kitazawa96}. These methods agree on a
critical point at $b_c(\delta_c) \simeq 0.6 (0.25)$. SE has been
applied to the dimerized spin$-1$ chain \cite{Singh88}. From the
SE of the ground state equal time structure factor and its second
moment they obtained $b_c \in [0.56, 0.64]$. While the latter type
of SE was based on ground state properties only, a SE for the
elementary excitations and the triplet gap has not been given to
our knowledge.
\begin{figure}[tb]
 \begin{center}
 \includegraphics[angle=270,width=0.48
 \textwidth]{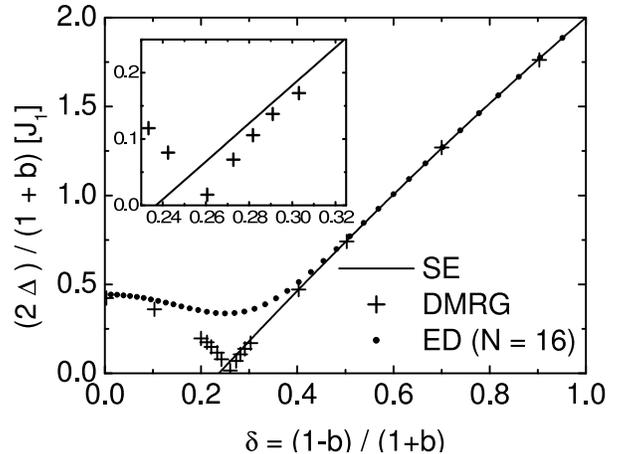}
 \end{center}
\caption{Solid line: integrated Dlog-Pad\'e $(4,3)$ approximant to
the $O(7)$ SE, Eq.(\ref{dispersion}), for the one-triplet gap
$\Delta \equiv \omega(k=0,b)$. Axes have been scaled to allow for
a comparison with other methods. Dots: ED on $N=16$ sites with PBC
in the dimerized spin$-1$ sector \cite{Brenig-tet}. Crosses: DMRG
results from \cite{Kato94}. Inset: zoom of the transition region.}
\label{gap}
\end{figure}

Fig.\ref{gap} shows the integrated $(4,3)$-Dlog-Pad\'e approximant
for the triplet gap from our SE at $O(7)$. As from
Eq.(\ref{dispersion}), the gap is located at $k=0,\pi$. The axes
have been scaled to allow for a comparison with two other
approaches, i.e. ED on $N=16$ sites with PBC in the dimerized
spin$-1$ chain sector \cite{Brenig-tet} and DMRG results from ref.
\cite{Kato94}. The latter two cover the complete range of
$b(\delta)\in [0,1] ([1,0])$.  For $b(\delta) \lesssim 0.4(\gtrsim
0.4)$, there is a very good agreement among all three approaches
displayed. For greater(smaller) values of $b(\delta)$ finite size
effects become evident in the ED data. The results from the SE
however remains very close to the DMRG over the complete range of
inter-tetrahedral couplings, up until the critical point. From the
inset of Fig.\ref{gap} one can observe that the SE slightly
over(under)estimate the critical value of $b_c(\delta_c)$ as
compared to the findings of DMRG.  In fact, performing standard
Dlog-Pad\'e error analysis, by evaluating the scatter of the
critical point $b_c(\delta_c)$ as obtained from different
approximants, we get $b_c(\delta_c) \in [0.612,\, 0.614] (\in
[0.239,\, 0.240])$,  whereas DMRG estimates $b_c(\delta_c) \in
[0.59,\, 0.61] (\in [0.24,\, 0.26])$ \cite{Kato94}. The small
difference may be due to non analytic corrections to a plain
power-law behavior of $\Delta$. Indeed, the gap has been claimed
to close as $\Delta \sim (\lambda)^{\frac{2}{3}}/|\log
\lambda|^{\frac{1}{2}}$, with $\lambda = |b-b_c|/b_c$, in the
vicinity of the critical point \cite{Schulz86,Totsuka95}.  In case
of such logarithmic corrections, deviations as those shown in the
inset of Fig.\ref{gap} are likely to occur and even higher order
SE would be required to improve the agreement very close the
critical point. Further evidence for the relevance of logarithmic
corrections stems from the critical exponent of $\nu \in [0.98,\,
0.99]$ which we extract from the Dlog-Pad\'e approximant which, as
can be seen also from Fig. \ref{gap}, varies almost linearly with
$b(\delta)$ close to the critical point. This is at variance with
extrapolation of ED data, which predict a critical
\emph{effective} exponent $\nu \in [0.7,\, 0.8]\,(\Delta \sim
\lambda^{\nu})$ \cite{Totsuka95}, and with the onset of additional
curvature which can be observed in the DMRG results close to
$b_c(\delta_c)$ in the inset of Fig.\ref{gap}. This issue should
be addressed in future studies.

\section{Conclusions}
To summarize, we have studied zero temperature properties of a
tetrahedral cluster spin chain using a series expansion technique
based on Wegner's flow equation method. Starting from the limit of
isolated tetrahedra, we have obtained expansions up to $O(7)$ in
the inter-tetrahedral couplings for the ground state energy and
the dispersion of the elementary one-triplet excitations within
the dimer phase.  The ground state energy has been used to
determine a first order quantum critical line which separates a
dimerized spin$-1$ chain sector from a singlet product phase. Our
findings are in excellent agreement with those of ED. A second
order critical line for a dimer-to-Haldane phase transition was
obtained by analyzing the closure of the triplet gap. Again, very
good agreement was found with the previous results from ED, DMRG
and QMC.

\section{Acknowledgments}
The authors take pleasure in thanking specially A. Honecker for fruitful
discussions and D. C. Cabra, F. Heidrich-Meisner
and G. Rossini for useful comments. This research was supported in part 
through DFG Grant No. BR 1084/2-2. Preparing this manuscript one of us (W.B.) has
benefitted from the hospitality of the KITP at UCSB
and partial support trough NSF Grant No. PHY99-07949.


\begin{thebibliography}{99}
\bibitem{electronic_data}
On appearance of this paper further results are available on
request from the authors. These results are too lengthy to be
displayed in written form but will be made available
electronically.
\end{thebibliography}
\end{document}